# What is wrong with MDPI: Is it a predator or a serious competitor?

Pasi Fränti

Machine Learning, School of Computing, University of Eastern Finland, Finland

**Abstract:** Hunt for predatory journals is based on binary division of the publishing world to legitime and predatory journals. This leads to difficulties in labeling publishers like MDPI with questionable practices. However, the root cause is not the publisher itself but the problems in the peer review system. These problems have created markets for the for-profit publishers shaking the understanding of conservative researchers and their belief about what is good science. In this paper, we analyze the problems via examples based on our own experience as author, reviewer, guest editor, associate editor, and editor-in-chief. We discuss how MDPI has addressed these problems.

## 1. Introduction

Ever since Jeffrey Beall [1] published his first list of predatory journals, the topic has gained the attention that it deserved. However, the publishing world has changed a lot since then and the single-minded question of predatory vs. legitime journal has evolved to something that is much more difficult to evaluate. There are several reasons for this. First, the quantity of research has increased a lot. Second, publishing has concentrated on major publishing houses gathering more journals, but also attracted new publishers like MDPI to compete for the money, fame, or whatever the reward. Third, the main problem is not really the capability to detect predatory journals but the problems of the peer review system.

Richtig [2] pointed out that it is important to recognize predatory publications because *common people reading these papers cannot easily recognize the validity of the journals*. This can be a problem in medical sciences where people reading fake results could lead to life-threatening problems. In computer science, however, the world is very different. Researchers keep proposing new methods at a fast pace and very few study the topics deeper. One reason is that the results tend to get outdated fast. Another reason is that people, by default, do not trust the published methods to work in practice as well as claimed, and researchers are likely to adopt only selected parts of them. New results are often easy to produce by computer simulations, even easier than providing fake results convincingly. Fake results are not a life-threatening issue; if a method does not work in practice, it will be forgotten fast.

Beall was the pioneer with his blacklist of predatory journals. Despite its shortcomings, the list was a useful tool for researchers until it was stopped. The list was later continued by others but nowadays, more efforts have been put on so-called whitelists of journals that are classified as not being predatory. However, hunting for predatory journals has become less relevant, counterproductive, and missing the root cause.

The real problems are in the peer review system which is hampered by long processing times, poor quality reviews, and editorial decisions based on secondary criteria. The main purpose of peer review should be to validate the results, because once published, the results start to circulate in the literature. But this original goal seems forgotten, and the review is focusing on secondary criteria such as is the article *good enough* (elitism), is the topic *relevant* (topic bias), is it *significant* (significance bias), does the *formatting* follows all the guidelines in detail (rule bias). Sometimes reviewers may just "*vote for rejection*" thinking that the paper was just a candidate in a competition.

Hunting fake papers and predatory publishers can help to clean the field but the publishing world is not black-and-white. The situation is demonstrated in Figure 1. In theory, we can distinct the legitime (good) journals from predatory journals by some quality measure and drawing a line between them (threshold). However, the reality is something else. Too liberal selection (threshold set too low) would miss a lot of predatory journals making the detection ineffective. Too conservative selection (threshold set too high) would cause many false detections causing protests by the legitime journals that were wrongly labelled as predatory. This was likely the reason why Beall was forced to shut down his list.

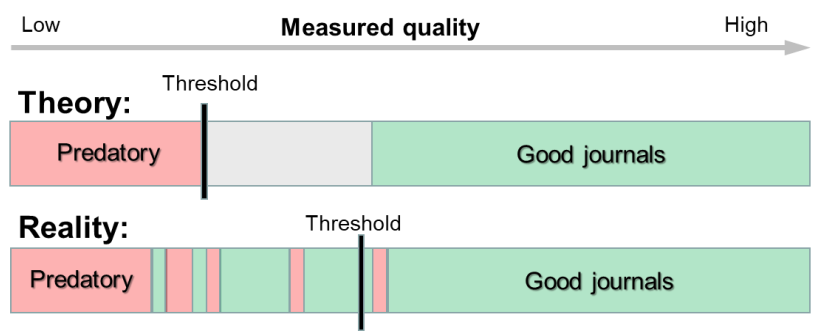

**Figure 1**. Classification of journals as predatory (watchlist) or legitime (safelist).

Labeling entire publishers like MDPI as predatory would not solve much but would just enhance the existing problems. The root cause is the problems in the peer review system and the rise of new publishers like MDPI is merely a consequence of this. The increasing amount of (valid) research results need to be published somewhere. If there is no space in the existing journals, new publishers will emerge to fill in the gap.

One of the reasons is that many journals focus on their reputation by elitist behavior. If legitime journals are ready to accept only the top of the crop, what to do with the rest? This elitist behavior creates a need, and this need is what publishers like MDPI are trying to utilize. The predatory journals will also get their share. However, labeling major publishers like MDPI entirely as predatory would have major consequences. The more there are unpublished papers in the market, the more there are predators as well. The poorly working peer review system pushes researchers (especially the younger ones with limited time for research) to seek easier alternatives.

The effect is demonstrated in Figure 2. Assume that there are 1000 papers of which 300 are valid and 700 non-valid. Assume also that a top journal (Journal 1) successfully rejects the 700 non-valid papers but accepts only 10% of all papers using some secondary criteria. The remaining 200 valid non-accepted papers are submitted to the 2$^{nd}$ choice with a higher (20%) acceptance rate. The process is then repeated until all valid papers have been accepted. The result is that even after five rounds, 34

(10%) of the valid papers are still not accepted by any journal. According to these theoretical calculations, the expected number of submission rounds to reach acceptance is 2.97.

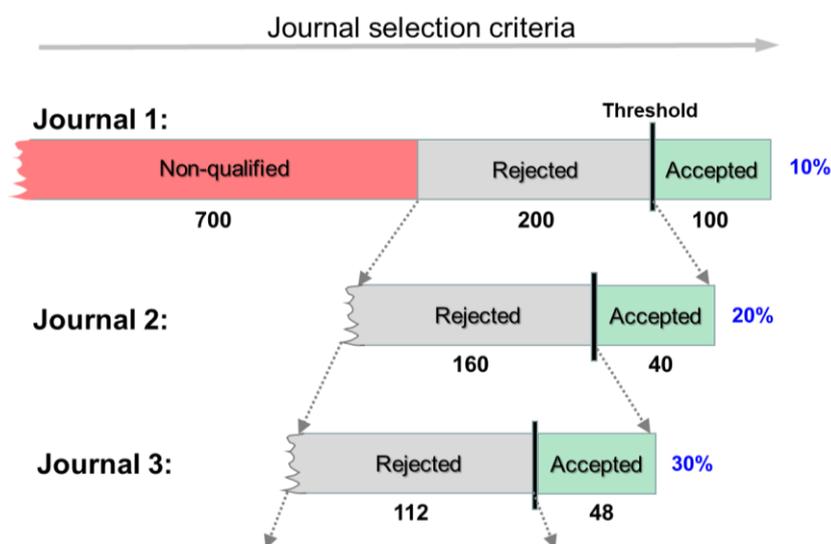

**Figure 2**. Effect of elitist selection of papers when the acceptance rate for the first journal is 10%, the second 20%, and so on. Assuming there are 1000 papers of which 300 are valid. The number of valid non-accepted papers decreases to 200, 160, 112, 67, 34, 13, 4, 1.

The rest of this paper is organized as follows. We first recall how the predatory journals have been detected, and what kind of quality criteria have been applied in literature. We discuss how these criteria match with the operating practices of MDPI and other (legitime) publishers.

We then focus on the problems of the peer review system. They have been known long. Smith [3] summarized it as "*slow, expensive, highly subjective, prone to bias, easily abused, and poor at detecting gross defects.*" Haffar et al. [4] provide a review of what kind of biases others have reported and concluded that "*it is time to improve the quality, transparency, and accountability*" of the peer review system. We discuss the problems from three perspectives:

- Efficiency
- Quality of peer review
- Special issue practices

These are not the only factors worth considering but the ones we identified. We provide concrete examples demonstrating how the problems appear in practice. The arguments used in this paper are supported by examples, logical arguments, and literature. All arguments are based on the author's long-term experience in scientific publishing, mostly as an author, but also as a reviewer, guest editor, associate editor, and editor-in-chief. This paper tries to be objective, but some bias might exist as the author evidenced many of the cases in person. Some examples might also appear surrealistic, but they all are real.

Our main argument is that the root cause is the problems in the peer review system rather than the journal being open access (becoming the norm anyway) or working on a for-profit basis like MDPI. Most academic publishing is made by large for-profit organizations anyway and non-profit is a minority [5]. The five largest publishing houses control more than 50% of the market and they are making huge profits according to [6]. What are the motives of the journals and their stakeholders, and how do they affect their operation and quality?

## 2. Predatory publishing

Predatory journals are ready to publish anything just to collect the publication fee. This raises problems because scientific publications are expected to be reliable with valid content and controlled by the peer review process *before* publishing. Plagiarized papers also appear frequently in predatory journals. According to [8], 68% of all the papers in predatory journals in nursing were plagiarized.

### 2.1. Detecting predatory journals

Detecting predatory behavior in journals has focused on indirect evidence. Early days people submitted randomly generated papers to predatory conferences to see whether the paper was accepted. This simple test would apply to detecting obvious predatory publishers as no legitime journal should publish nonsensical papers, but this activity seems not widely used.

Bohannon [9] submitted a fake scientific paper to journals listed by Beall (watchlist) and DOAJ (safelist). Predatory journals [1] were expected to accept the paper, and legitime journals (DOAJ) should have rejected it since DOAJ is a safelist of open access publications despite the attitude towards open access publications was highly suspicious around 2013.

The results in Figure 3 show that 84% of the papers in Beall's list were accepted (among those for which the decision was reached). The rest 16% that rejected the paper could be either legitime journals or predatory with some efforts to pretend to be legitime as they passed Bohannon's test. Nevertheless, this experiment gave rough evidence of the accuracy of Beall's list. The fact that 54% of DOAJ journals accepted the fake paper showed that, around 2013, DOAJ failed to serve its purpose as a safelist. Nowadays the result might be very different.

In literature, open access has been commonly associated with predatory publishing [10] but the attitude towards open access publication has changed a lot. Björk and Solomon [11] compared the impact (citations) of open access and subscription journals and showed that after controlling for discipline, the age of the journal, and the location of the publisher, the differences in impact largely disappeared. In fact, the situation can be the opposite because open access publications have a citation advantage over subscription journals [12]. This is simply because they are freely available to a wider audience.

In brief, a predatory journal is open access for practical reasons because it would be very hard to sell it to libraries as a subscription journal. Journal being open access, however, does not serve much of evidence of the journal being legitime or predatory.

Zernes [13] suggested identifying types of unethical behavior in any journal rather than trying to distinguish between good and bad journals. This is a good and sensible suggestion. Safelists and watchlists are still needed to help researchers to recognize obvious frauds but having good tools to evaluate the quality of articles would be much more important. Even classification to valid and potentially problematic papers, or some quality score, could help. Evaluating journals would then become simply statistics based on the quality indicators of the individual papers.

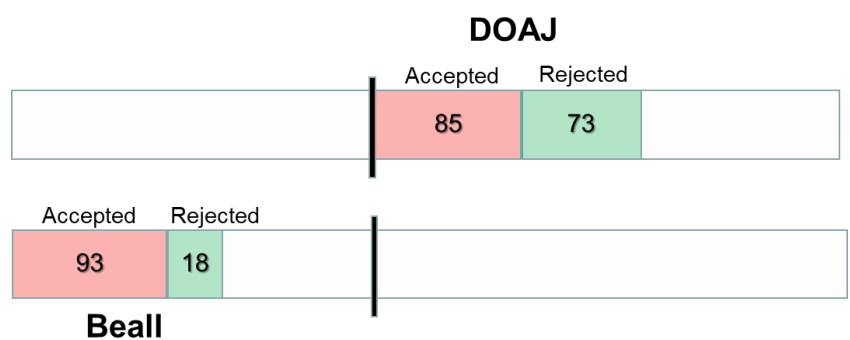

**Figure 3**. The number of journals that accepted and rejected the fake paper submitted by Bohannon [9]. Papers in Beall's list are assumed to be predatory (watchlist) and expected to be accepted while DOAJ (safelist) is supposed to represent legitimate journals and reject the paper accordingly.

**2.2. Watchlists and safelists**

Maintaining *watchlists* (blacklist) and *safelists* (whitelist) is a tedious task as new journals emerge fast. Koerber [14] pointed out the obvious that any attempt to maintain a safelist or watchlist is not perfect and will become outdated fast. Problems are false classification to predatory and journals that do not appear in any list. Current watchlists include *Beall*, *Cabell's predatory report*, *Dolos list*, and *Stop Predatory Journals*. Apart from Koerber's report, we are not aware of other efforts to report the accuracy of the lists.

Several countries, including Finland[1], maintain its own journal ratings which effectively implements a safelist. The process of updating the list is a bit slow, depends on the researchers' own initiative, and has some bias, but the result is useful as it provides a trustworthy safelist. The exact journal rating might be questionable in some cases, but the usefulness comes from the fact that one can easily verify many journals as being valid by reading that list. Problems are caused by journals in which no Finnish researcher has published and has not yet been promoted to the list.

Chinese Academy of Sciences also published its own watchlist as the *Early Warning List of International Journals* with only 65 journals but including major titles like several MDPI, IEEE Access, and four Hindawi journals. They used criteria like (1) retractions, (2) self-citations, (3) rejection rate, (4) article publication charge (APC). According to Petrou [15], the list has already had a significant impact by reducing the number of submissions to MDPI journals and IEEE Access by Chinese authors. He stated that the common denominator of the journals in the list is not their predatory nature but the fact that they attract considerable APC payments from China. In other words, the motivation of the list seems political; to prevent payments from flowing out from China. Petrou also criticized the last two criteria (rejection rate and APC) as being outdated, and that the self-citation relates more to citation indicators like impact factor than the predatory nature of the journal.

The classifying of journals has also moved towards classifying publishers. It is indeed easy to outsource the quality control to the publishers and rate the entire publishers such as ACM, IEEE, Elsevier, and Springer. Problems arise with publishers like MDPI which have journals both with ratings 1 (qualified) and 0 (not qualified) in Finland's list. Many researchers have already published in MDPI journals. Should we consider those papers also disqualified as well if the entire publisher were put on a watchlist (0 rating)?

---

[1] https://www.tsv.fi/julkaisufoorumi/haku.php?lang=en

**Precision & recall:** The binary classification of journals being predatory has also a big limitation as it focuses only on one side of the coin: the *false acceptance* of flawed and fake papers. The other side of the coin is the *false rejections* of valid papers. This is what drives researchers towards dubious journals and causes the problem in the first place.

In computer science, the evaluation of a system with two contradicting goals is known as two-objective optimization and can be measured by *precision* and *recall*. Suppose we have some measure to estimate the quality of the journal and need to draw a line (*threshold*) between the legitime (above the threshold) and predatory journals (below the threshold) as shown in Figure 1. Precision measures how many (%) of the detected journals are truly predatory and recall how many predatory (%) were caught. Using just one measure we can achieve high recall (detecting most predatory journals) simply by setting the threshold very high but at the cost of low precision (many false detections).

## 2.3. Quality criteria

Detection of predatory journals is typically made via indirect information such as impact factor, fake address, fake editorial board, general emails (Yahoo, Gmail) which are easy indicators but not conclusive. Narimani and Dadkhah [16] presented seven criteria: emails not available (or are general), publication charges, journal names resemble legitime journals, fake impact factors, aggressive advertising (spamming), unclear process, lots of papers per issue and outside of the scope. Shamseer et al. [17] found out that 66% of predatory journals homepages contained spelling errors, 63% used unauthorized images, 33% used bogus impact metric. The same numbers for open access journals were 6%, 5% and 3%.

Cobey et al. [18] extracted 109 characteristics from 38 articles. Most focus was on journal operations and the most common factors were lack of transparency (19 statements), poor quality standards (17), and unethical research or publication practices (14). However, many highly reputed legitime journals suffer from the same above-mentioned problems and many quality indicators apply to them as well. For example, lack of transparency is common with publishers like IEEE who do not even publish the name of associate editor effectively making the decision. Open access with publication fees is also normal practice for major publishers as well. Suffering from poor editorial work, their main advantage remains just their brand name. More attention should be paid to the articles published in the journals instead of secondary criteria and hunting for predatory journals.

**2.4. MDPI practices**

Oviedo-García [19] reviewed a wide variety of definitions for predatory journals and studied how well MDPI journals met the selected criteria. This research question takes the binary classification approach (legitimate journal versus predatory journal) to the level of an entire publisher (MDPI). The obvious answer should be that MDPI is not a predatory publisher as it does not fulfill the main criteria, which is to operate solely to gain money without any scientific review. Predators have the extreme behavior of accepting as much as possible to maximize revenue whereas the suspicion with MDPI can be at most how much the publisher prioritizes self-interest at the expense of scholarship.

One observation of questionable practice of MDPI in [20] was a higher proportion of self-citations (within the same publisher). However, self-boosting itself is typical for many journals and could be argued by the need to show the submitted paper to have relevance to the journal roughly along the following lines (from a reputed journal):

> "*Please make sure the revised version is relevant to the readership of the Journal of Pattern Recognition. Make sure you cite **recent** work from the field of pattern recognition that will be relevant to our readership.*"

The request is subtle and harmless, but the intention is clear. According to Davis [20], it is routine practice for editorial boards to think about strategies to increase their impact factor scores, and it *normalizes unethical behavior*. Could it be that MDPI is doing this just more aggressively? Typical predatory journals act as robots and would not even care about impact factor as they are just for the money. Self-boosting is a human characteristic; robbers would not bother. If self-citations are considered a problem, they can be eliminated from the citation counts. It becomes questionable if applied at the publisher level, and especially if it affects the acceptance decisions.

Impact factor itself has lost its original significance as a measure of journal reputation but still serves a purpose to a certain extent; see [6]. Björk et al. [7] showed that papers in legitime journals were cited 18.1 times during five years (9% of papers were not cited at all), compared to 2.6 citations of papers in predatory journals (56% were not cited at all). The difference is a factor of 7. It would be interesting to know the impact factors of most common predatory journals.

Another alerting factor Oviedo-Garcia [19] pointed out was the size of the editorial board. MDPI journal *Applied Sciences* has 24 editors and 2165 editorial board members, which seems huge at first. The corresponding numbers of a reputed journal, *Pattern Recognition*, has 9 associate editors-in-chief and 234 associate editors with a much narrower focus. The journal receives about 3000 submissions per year and, according to EiC, the journal is struggling with the volume. The journal website claims an average of 13.8 weeks for the first decision, but according to our experience, it takes already 2 months just to get the review process started indicating the number is not trustworthy. The need for a large editorial board seems well-motivated.

**2.5. Evaluate papers – not journals**

I would suggest taking the evaluation directly to the level of individual papers. Why do we even need to measure the quality of journals? Is it only because we are not able to measure the quality of the publications in it? The quality of a journal can then simply be concluded from its papers. The remaining factors at the journal and publisher levels would be their efficiency and how well they serve the authors during and after the submission process.

Impact factor is an obvious measure, but it can be calculated only after the publication by citation counts. Flawless methodology and correctness of the results would be other relevant criteria worth measuring, but they require expert evaluation.

**2.6. What is worth publishing?**

Everything is not worth publishing but can we really decide, and should we even care? Are you the qualified scientist to decide whose (valid) work is worth publishing? What about the scientists having inferior skills compared to you? People tend to respect (or envy) the ones recognized to have higher skills and merits and belittle others.

It is a kind of elitism. According to [21], the editor-in-chief, Buckley, had been working hard to boost the visibility and reputation of the MDPI journal *Nutrients* but became worried about a sharp rise in the number of low-quality submissions. To solve the problem, Buckley was reported to aim at increasing the rejection rate of the journal from about 55% to 60-70% causing a clash between the editor and the publisher. MDPI CEO at the time, Franck Vazquez, objected this arguing that submissions must be evaluated based on quality, not by "*setting an artificial rejection rate*", and "*if more papers are good enough, more should be published*".

It is easy to understand the editor wanting to build a highly reputed journal, but even easier to agree with the CEO's arguments: whenever an article is sound, it should be published, even when the novelty might be limited. If the paper is flawless and contains some useful results, then why it should be rejected from publication? Even into extending cases when the paper is merely re-confirming earlier findings. Journals may have different quality criteria but if a paper qualifies to these criteria, should we reject it because of the lesser expected impact? Aiming at a high rejection rate is elitism.

## 3. Efficiency

Editorial efficiency depends on the complexity of the system, and the efficiency of the individuals involved in the system. The paper may lay down first on the editor's desk, then on the reviewer's desk for weeks (or even months) without any action. Every person in the loop is a potential bottleneck, and the more in the loop, the longer the expected delay. If nobody manages the process, delays are likely to happen. Editorial systems may have automated email reminders, but many still rely merely on the editor's activity. Some journals may not even give any deadlines, or the allowed time is too long. Reviewers can work simultaneously, but otherwise the process is sequential, and delays tend to cumulate.

According to [22] the expected time for the first decision should be no longer than 1-2 months, adding 1-2 months for each revision, summing up to 8 months (assuming three revision rounds). This might reflect the (in)efficiency of a typical journal but there is high variation. MDPI has a much faster review process; 19 days to the first decision (2018 and 2019) and 39 days to publication [19]. Petrou [23] reported that the turnaround time of the ten largest publishers from 2011/12 to 2019/2020 decreased by 10% but it was mainly due to MDPI; the turnaround time (from submission to publish) of all other publishers remains static.

The difference is remarkable, and most of this originates from efficiency. Making a process efficient is not magic. It is professionalism. Very long turnaround times have been common and waiting for the review result for 12 months is not even rare according to my experience. My record is three years without even a single review report at the end [24]. The paper was accepted based on our initiative and self-organized review [**Appendix A; Case 1**].

Another paper took even longer [25]. This time we re-submitted (and published) the paper elsewhere without waiting (or even informing) the original journal. After four years, the journal woke up and reported that a significant part of the manuscript was found on the web and cannot therefore be accepted. It was the pre-print of the published paper stored in our repository [**Appendix A; Case 2**].

While these two are extreme examples, many journals in our field are very inefficient and waiting for one year for the first decision is not just an exception. This becomes very unmotivating for the authors and counterproductive when the result is a rejection with secondary criteria. It is a relevant question for the authors whether you want to risk wasting one year or a few weeks in the *review lottery* in the hope of getting a brand label for your publication.

So where does this inefficiency come from? We next analyze a few of the problems of the review process, discuss how MDPI has addressed these, and how the for-profit factor might affect this.

### 3.1. Finding reviewers

It is notoriously hard to find reviewers - good reviewers are even harder. The first challenge is that one must pinpoint interested readers and contact them one by one. After the publication, it is different. When the paper becomes widely available, interested readers will find it via web searches and recommendation systems. But the reviews are not needed anymore then. Instead, the paper will often be reviewed by people who may not even know the topic very well, and therefore, not very motivated. A good *recommendation system* would help.

One reason for the difficulty is that reviewing is not the core business of anyone. It takes time and reading (often immature student) papers without any direct reward can be unmotivating. Providing good feedback (especially for immature student papers) equals acting as supervisor. If you have no time or interest to give enough guidance to your own students, how likely you would spend time guiding someone else's students? MDPI applies vouchers that aim to address this problem.

It is not only to find potential names, but also to predict how many of the invitees will agree and eventually provide a review. The probability of agreeing is surprisingly low especially if the editor does not know the reviewer in person. One might need to send 10 invitations to guarantee 2-3 reviews (the typical lower bound). My record is about 25. Many invitees do not even respond. For the sake of efficiency, one should *overbook* and send more invitations than the required minimum. Not sending new invitations before the previous ones have been resolved (either declined or being idle too long) can significantly slow down the process.

Airlines overbook their seats all the time with the risk that some travelers do not get a seat, but the airlines have ways to deal with it. Why should journals not do the same? The risk is to occasionally have more reviews than needed. If this happens, the editor can simply wait until the review deadline and then cancel the pending invitations. It is highly possible that the reviewer has not even started to read the paper yet. Sending a reminder a few days earlier with a clear deadline is fine when the review is overdue (especially if agreed with the reviewer). Deadline extensions can be applied if considered important. This all requires just a well-organized editorial process where most routines are automated or handled by the staff.

MDPI finds potential reviewers efficiently through their **in-house team** with the help of a large researcher database. With good topic matching and recommendation system the search has become routine. The staff also takes care of the invitations efficiently. Reviewers are not left much time to react, and the review requests are aggressively followed up making sure they will not slow down the process by staying idle. This can be annoying at first but once you get used to it and realize its positive effect on efficiency, you might even start to appreciate it.

Most editors fail to do this part even remotely as efficiently as the MDPI staff. Either you need to be well networked to know lots of researchers willing to help you; or know the topic particularly well to be able to identify potential reviewers with a high topic match. Simple keyword searches from the journal database may not be effective and manually fetching authors from the reference list easily becomes a burden as most systems do not provide any automated tools to help this. A good recommendation system can provide significant help in this part.

**3.2. Doing the review**

The reviewer may not even have started to read the paper when receiving the first reminder. Deadlines do not need to be hard, but active (yet not too aggressive) chasing the review reports by reminders gives a signal that being timely is important. The reviewer may even recognize it as a good practice and consider later submitting his own work to the same journal.

MDPI operates this step significantly more efficiently than many others. First, reviewers do not need months to do the review. A good and thorough reviewer might benefit from longer review time but how often do you receive thorough review seriously overtime? One characteristic of a good reviewer is planning his time and not letting various duties chaotically pile up leading to quick-and-dirty overdue reviews. If the reviewer cannot allocate the time soon when the invitation comes, then there is a high risk he will not allocate enough time for it later.

There is no need to accept the "*Ok, I will try to find time for it*" attitude. The challenge is to find reviewers who know the topic and have enough interest to read the paper **in the near future**. MDPI journals give tight, one or two weeks, deadlines for the review which may sound unrealistic. But this is where you must decide: do you have time *within the next 7 days*, or not. If the paper is interesting, I usually read it soon anyway, often even in the same or the very next day mainly out of curiosity. And then process the review fast to prevent it from becoming a burden on my task list.

I may regret it later if found out the paper is of low quality. This is a dilemma. Researchers are motivated to read interesting papers but highly unmotivated to read poor quality immature papers. It can easily become a burden and takes more time to write a proper review instead of the planned, say 30 or 60 minutes (unless you are the killer reviewer who rejects the paper with excuses). One week is enough for reviewing most papers, and to detect obvious flaws. Papers that I held for two months are those that I have forgotten and buried on my desktop, and nobody reminds me about them. If you want to reject the paper, it does not take two months to write down your arguments. Delaying without a reason is simply unethical and disrespectful.

In the case of MDPI, the *initial* decision of the editor seems important. Once the editor accepts the paper to continue the review process, the system is built so that the acceptance is the default final decision. The in-house team seems also eager to push towards this goal. This strategy gives the authors lots of chances to revise, which may become a burden for the editor. The time given for the revisions is also short, 1-2 weeks, which adds to the efficiency of the process. It may not be enough for all major revisions but assures that most papers will be completed within 1-2 months if the authors work hard for it.

### 3.3. Efficiency of MDPI

The efficiency of MDPI journals comes from several factors:

1. Very fast and systematic reviewer search implemented by their in-house team;
2. Giving a very short time for the review like one or two weeks;
3. Active chasing the reviewers and editors to meet the deadlines;
4. Waiting nobody;
5. Minimizing the involvement of the editor;

This efficiency goes as far as ignoring a reviewer or even editor if he becomes idle. The journals have large editorial boards which are also aggressively used. If the original editor cannot take care of his duty in a reasonable time, another can be called as a replacement.

Ignoring the editors may sound harsh but why should we tolerate the slackness when the life of the editor has been made so easy even by removing the need to search for the reviewers and chase their reports? I have also been asked (as editor) to make decisions for papers I have never seen before. Sometimes I even knew the original editor and concluded that he did not take the duty seriously. The efficiency-driven operation originates from the founder of MDPI, Dr. Shu-Kun Lin, who was quoted as saying "*Don't make the scientists wait and don't waste their time by unnecessarily delaying a response to them; this is the difference that we want to make*." [26].

The questionable part is that the editors are involved only in the initial and final stages of the process. The staff sends the review invitations and the result back to the authors for revision without consulting the editor, who is left only to confirm the final acceptance when the staff thinks the revisions are completed; see the comment of Joseph Trögl in [27].

The efficiency is likely to have consequences on the quality of the overall process, but it effectively avoids delays. If one wants to understand the reasons behind these practices, it is important to notice that the founder and most daily operations of the company are Chinese. Those who have worked with Chinese (researchers or students) might be aware of the Chinese mentality and how it differs from that of Western thinking.

Chinese mentality enhances efficiency and flexibility. They are the main advantages of MDPI as well. The drawback is that flexibility tends to lead to cutting corners at places where people may not expect it to happen. Even if it would not jeopardize the quality of the process, western people often dislike it, and it affects their attitude toward the journal and publisher. The reasons are cultural differences which have a dramatic effect on motivation when Western people work under Chinese management; see [28].

In my opinion, there is also a high demand for publication forums suitable for Chinese researchers. Chinese students are notorious for their poor English writing skills. Despite of this, I have often heard a Chinese student saying how well he *understood* a paper written by fellow Chinese (in English); even if I would consider the writing of low quality. What would serve better Chinese scholars than an international journal operated by other Chinese who share the same mentality and use the same writing style.

## 3.4. Is non-profit required?

Beall [1] assumed that an open access journal is a predatory publisher by default, but this is not the case nowadays and all large publishers have adopted this business model in addition to the normal subscription-based model. The question is merely how the open access model influences the publication world.

First, operating for-profit has consequences but not as simple as money would mean an immediate loss of ethics and scientific principles. One important consequence is that for-profit creates motivation to work efficiently. The key operators in traditional journals are editors and reviewers who focus (or are assumed to focus) on quality control but they may lack motivation for daily operations, which is left to the staff. The journal staff is usually paid but their number can be small or even non-existent due to a low budget. This leads to inefficiency.

Some of the resource problems have been lessened by the tendency of journals to move under the umbrellas of larger publishers which can provide better editorial systems, recommendation tools, improved reviewer search, and more human resources for running the processes smoothly. At least in theory. It seems that publishers like MDPI have been able to fully utilize the advantages of a bigger publishing house and tune their processes also for efficiency. MDPI has large and shared staff taking over most routine tasks. There are plenty of editorial staff to look after the process and the operation never relies on a single person.

If being non-profit would be the main factor to label a journal as legitime, then one should scrutinize all journals. Do they have high scientific standards and ethics? Most journals are published by major commercial publishers, and they all make huge profits [6]. They have increasingly expanded also to open access publishing, and not because of equality or other noble goals, but simply to keep their business highly profitable [30].

One potential solution is the *diamond* (also known as *platinum*) open access model where neither the author nor the reader needs to pay. The digital publishing era has made this model economically more realistic as the most significant burden, the printing cost, has diminished. Some action in this direction is going on.² Another possible solution is so-called *Overlay* publication model, where the journal takes care only of the peer review process, and the actual publishing uses archiving services like ArXiv. This would reduce the publication cost to about 37 USD per paper; 1% of that of a commercial publisher [31].

The publication fee itself can be an indicator of a journal being predatory. According to Memon [29], the fees of predatory journals varied from 25 to 1800 USD, the median was about 160 USD. Shamseer et al. [17] had very similar observations with a median of 100 USD. Open access fees for legitime journals of large international publishers are between 1500 and 3000 euros according to [32]. According to the same source, Finnish institutions paid 1,639 EUR in open access payments in 2019, on average. We can conclude that legitime open access journals have either relatively high publicati nmhk on costs, or no cost at all whereas most predatory journals have a very low publication fee. According to my experience, authors are more concerned about the indexing, journal ranking, impact factors and whatever measures their universities are applying to evaluate their researchers.

---

² https://www.scienceeurope.org/our-resources/action-plan-for-diamond-open-access/ and https://diamasproject.eu/

## 4. Quality of the peer review

The downside of the faster speed is that it may compromise quality [15]. Crosetto [21] argues that fast turnaround is good for science **only if** the quality of the peer review can be kept up. One might expect that the reviewers are the gate keepers of the quality control. However, this is rarely the case.

### 4.1. Long review time

A longer review process may help to improve quality as it gives the authors more time to discover (and fix) problems. Having a few weeks' break from the paper will help to see it with a fresh mind which helps to spot errors. A pedantic attitude helps in this as no suspicion is left unanswered. In theory, more review rounds help to reduce the remaining errors and a shorter review cycle may reduce some of this self-correction mechanism.

However, according to my experiences critical errors were almost always spotted by us, the authors, and rarely by the reviewers. Time helps in this, but a very long waiting time easily becomes counter-productive. If you do not get the feedback fast, you start to forget the details and eventually lose motivation. If the reviewers have nothing more to say than complain about irrelevant details caused by hasty reading, or merely find excuses for rejection, there is no reason to wait for months.

For example, the paper [33] was in the review process for four years. One journal spent 12 months without any useful review reports. The other three were more efficient but the quality remained poor. The only relevant feedback was obtained by the IEEE vice president after a rebuttal (**Appendix A; Case 3**) and only to improve clarity. Validity was never questioned, and the main contribution remained the same during the process. Long review periods are common even in reputable journals.

Have you ever seen an interesting paper on ArXiv.org without a peer reviewed journal version published later? Did it make you wonder whether there is some flaw in the paper, or whether the authors just lost motivation and gave-up because of poor peer review experience? Some of these papers are even highly cited, and some have potentially useful results. In contrast, IEEE publications are characterized by complex engineering solutions which often have no practical usefulness. Complex engineering solutions are preferred over the simplicity and usefulness of the ideas. The key to getting a paper accepted relates more to *playing the game* than its usefulness.

### 4.2. Review reports

In contrast to real quality control, reviewers often pick up irrelevant details just because they dislike the paper due to the choice of topic or methodology. The attitude of many reviewers can be very narrow-minded, like accepting only papers using the same research paradigm or methodology that the reviewer himself uses. Competing results can be treated harshly. Many reviewers seem also totally incompetent for the task. Making good research requires different skills than making constructive review reports.

Typical reviews in my field are anonymous (single blind) although double-blind is also used. Squazzoni et al. [34] encouraged journals to test different peer review models for better transparency and how to best engage and **reward** the reviewers. Open review is one such possibility. Bravo et al. [35] studied the effect of publishing peer review reports (with name or anonymous). It did not affect turn-around time but decreased the willingness of more senior academic professors to review. Those who agreed to publish their names (only 8%) gave a more positive and objective review.

The good and helpful reviews are rare and a matter of luck. According to my experience, it used to positively correlate with the quality of the journal. There was a good chance of getting at least one useful comment from a quality journal whereas the chance from a lower quality journal was close to zero. This advantage seems to have diminished and elitist reviews have become the norm with "*not important enough for the journal*" phrases.

Reject recommendation is often given merely for convenience. If your recommendation was to revise, you would likely be later asked to review the revised version. One reason why an unmotivated reviewer recommends rejection is hoping not to see the paper again. The other extreme is the *nice guy* reviewers according to whom everything is acceptable with at most some minor modification.

When it comes to MDPI, the problems of the review report quality are the same as with other journals. The differences are mainly a consequence of the operation practices. First, the use of rewards and a significantly faster review cycle is likely to attract more nice guy reviewers. The last time I faced misleading claims was a paper that presented presumably a real system, but the documentation was filled with concepts and methodologies never used in the system. Their main purpose seemed to make it look more convincing. Other reviewers were nice guy reviewers who seemed not bothered by this.

The second difference is the reputation of MDPI. There is an invisible "*we are not too picky*" sign on the top of MDPI. Reviewers care less about what goes into these journals; see the comment section of [27]:

> *I've published in an MDPI journal and reviewed at a few – most reviews are not very rigorous and I personally don't put a lot of effort in when reviewing for MDPI because I know that even if I recommend reject it will come back for revisions.*

The attitude of the same reviewers can be completely different if it were for a reputed journal.

**4.3. Elitism**

A paper I recently reviewed for IEEE Journal was rejected with the words: "… *I regret to inform you that your paper cannot be accepted. The journal is very competitive, and we only accept approximately 10% of the submitted papers. We regret that we do not have space for some quite good papers.*" In other words, the journal openly admits that it applies elitism. The paper was reviewed also by its content, but this is not always the case. Desk rejections based on secondary criteria are common.

The stricter elitist selection can theoretically decrease the number of flawed papers in the system but only in the case of the elitist journal. Rejected papers will be submitted elsewhere, both the valid and the flawed papers. If a paper is rejected without any content review, it does not help the quality control as the paper continues circulating as such without any error fixing. It is just the tip of the iceberg that gets accepted, say 10%.

What about MDPI papers, do they have significantly more papers with flawed content? This I cannot answer. It would be interesting to see a detailed evaluation of how the quality of papers differs between MDPI and other publishers. In the case of my research areas, I have not recognized any clear pattern in which journals to follow, or which to ignore. Interesting papers can appear anywhere and are not limited to the top journals. The overall quality seems lower in MDPI journals, and important papers are less likely to appear there.

The main contribution of MDPI is its effect on publication markets: significantly faster review process with less elitist attitude, which both have been highly welcomed. This hopefully pushes other journals to improve their operating practices. The downside is the questionable reputation which prevents many from submitting to MDPI journals at all. This can go to the extent that some would label the entire career of a researcher void if published even a single article in an MDPI journal; see Figure 4. A threat of MDPI is that it may diversify the review quality further from one extreme (elitism) to the other (everything is accepted).

**4.4. Review bias**

As a result of elitism, it is likely that the published papers still contain errors. Papers rejected by secondary criteria indicate a higher possibility of the papers having content errors as well. However, when the rejections start to be the norm and the reasons are only secondary criteria, the system will not work. Typical secondary criteria are:

- The paper is not important enough
- Not fit to the scope of the journal

I call these *garbage reviews*. If they do not contain any useful information on the content, it merely drives the authors to seek shortcuts rather than improving the quality. Especially since the process often takes long, it is no wonder that publishers like MDPI have appeared. The difference between MDPI and reputed journals in this regard can be just the *time* to get the garbage reviews. Reputed journals are likely to have more expert reviewers but according to our experience, this does not show as a better review quality. Traditional journals are just *slower* to provide garbage reviews.

One motivation for providing garbage is jealousy. Another is that the reviewer dislikes the paper for some reason. Jealousy is harder to deal with as the reviewer is motivated to invent seemingly convincing excuses, or to use secondary reasons for rejection. It will never stop no matter what revisions you do. One of our papers [36] was originally rejected (from a MDPI journal) because of a hostile reviewer motivated by jealousy. The reviewer kept on providing garbage reviews revision after revision; see [**Appendix B; Case 1**]. In the end, the editor concluded that "*There **must be** something wrong with the paper*". Even a theoretically good paper like [37] can get rejected for secondary reasons if the results contradict earlier findings.

A third reason for providing garbage reviews is laziness. The reviewer agrees to review but does not wish to deal with the paper due to lack of time, skills, patience, or knowledge on the topic. A simple shortcut is to find a few excuses or use secondary reasons for rejection. I will next summarize the types of bias we have experienced during the (mostly single-blind) peer review process. They can be unconscious or used as excuses.

**Rule-bias:** Journals usually provide a template to guide the writing, with long and detailed instructions. Many of these guidelines are insignificant for the review and become meaningful only in the final publication phase. However, they can provide endless resources also for the reviewers to find excuses for a rejection. Grammatical errors are the most common scapegoat. Even professional proof-reading does not prevent "*The English should be reviewed*" type of comments.

Page overlength can also cause rejections. A reputed journal like *Pattern Recognition* provided such rejection after **2 months** of process without any comment on the content. Another paper for the same journal was rejected because of being too short even if the length was within the limits - it just had different line spacing making it appear too short. It took 10 seconds for the authors to fix but 3,5 months (1,5 + 2 months) in total just to get the paper into the review process. The result was two useless garbage reviews.

The formatting rules can become a burden when the target journal changes, and the paper requires excessive re-formatting. I have even witnessed a student deciding to submit to a predatory journal (unknowingly) merely because the formatting rules of the intended journal were too tedious to handle, and the rules of the alternative (predatory) journal were much more relaxed. Garbage reviews can also include a suggestion to use their favorite tool (namely LaTex). Many reviewers have a negative bias for papers not written by LaTex.

**Significance bias:** Researchers tend to be very narrow-minded when it comes to the topic. They often consider their research very important and that of others insignificant. The work of others is considered merely re-inventing the wheel, or an obvious application of some existing scientific fact like the rule of gravity. Studies confirming earlier results are considered "*widely known*" and those contradicting "*flawed*". For a paper with a clear simple idea, a common reaction is: "*It is too simple; it cannot be novel*".

This bias leads authors to present their results in an overly complex manner using fancy terminology and trying to hide the real (minor) contribution. No matter how clever the idea, it is risky to present it in a simple manner. This applies especially well to engineering journals like IEEE. Many good research results seem obvious only after their discovery, almost even trivial. The best ideas are simple.

The importance of the results is highly subjective and will become apparent only after the peer review process. The best results will be widely adopted (and cited) while other results may still have their role. Insignificant results will be quietly forgotten. In the review process, these factors are paid too much attention.

**Relevance bias:** Relevance to science is a valid criterion but relevance to a particular journal is a secondary criterion. The idea of having journals with a narrow focus trying to concentrate on all the relevant papers in the same area is idealistic; it will not happen. It will just cultivate elitism and papers left out need to be published elsewhere. In clustering, the papers are scattered all over and the relevant papers can be anywhere. It does not matter what is the title of the journal, as long as the paper will pop-up on the computer screen because of keyword search or by recommendation system. The title of the journal does not matter. The papers matter. The special issue practice demonstrates the relevance bias well; see the two cases in [**Appendix A; Cases 3 and 6**]

**4.5. Validity**

Validity in computer science refers to the correctness of the algorithms, computer programs, and the relevance of the data and methodology used in the experiments. Running the experiments is relatively easy. A consequence is that fake results are rare. It is simply easier to achieve some real progress than to provide fake results convincingly.

Lots of research is done by students with inferior writing skills especially when the supervising professor does not contribute to it. Problems with the paper relate to re-producibility, missing the state-of-the-art, and low-quality documentation. Obvious flaws appear but are less common. Even poorly prepared student papers may have a valid result. It is not the bottom of the barrel we need to worry so much about, but what is in the middle.

It is a common tendency of authors to present their results in over-positive light. This can go as far as purposely omitting the state-of-the-art. Selective use of data is common to show the new results in a favorable light. Younger researchers may not even be interested in digging deeper and tend to be over-optimistic about their inventions. Dealing with such papers may not be pleasant but not because of the lack of validity.

I try to follow one principle: if I learn something from the paper, I should not recommend rejection. I mean something that I would not have learned by reading some schoolbook material. Exceptions are when the technical quality of the paper is too low. And even when recommending rejection, I try to provide some constructive feedback on how to improve. The more effort the authors have put in, the more detailed comments I should provide.

The biggest problem I face as a reviewer is the poor writing quality. I cannot understand the method or the results without excessive focus. The second biggest problem is the authors do not know the state-of-the-art. In computer science, this is common. It is difficult to know everything that is happening in the wide and fast-evolving research. Researchers get new ideas all the time and implement so-called *epsilon improvements* with the help of the many software libraries.

Nevertheless, a lot of papers with valid results exist and they need to be published. Most of them would benefit from much better writing and more appropriate literature coverage. But the authors seek reward for their work, even if it were just epsilon research. If there is no room in the legitimate journals, these works have a high chance to feed the predatory journals. To sum up, science would be better served by improving the operations of valid journals than by hunting for the predators. Revealing predatory journals is important but it should not be the focus, and not become a witch hunt.

## 5. Special issue practices

One of the questionable strategies of MDPI is to attract submissions via special issues whose topics are quite wide. It works as human nature is appealed by "*special offers*". The publisher is also very active in inviting people to act as guest editors and going as far as proposing preliminary titles for the special issue to make it easy for potential editors to accept the invitation. But MDPI is not the only publisher using this model excessively; MDPI, Frontiers and Hindawi all had more than 50% of their papers due to the special issue practice [23]. This is just a marketing strategy to attract more papers. It seems common for journals that depend on article processing charges Hanson [38].

People in general dislike this aggressive marketing via special issues [39]. Crosetto [27] analyzed this practice further by showing that the number of MDPI publications grew from 36,000 (2017) to 167,000 (2020) based on two things: a lot of special issues and a very fast turnaround. The turnaround times of papers halved from 2016 to 2020, whereas the number of special issue papers increased remarkably to 7.5 times. Crosetto agreed that the fast turnaround itself is good and clearly an asset of MDPI.

Regarding the growth of the special issues, Crosetto concluded that MDPI is likely to shift towards *more predatory* over time. It indeed seems that MDPI has been overexploiting the markets and in this way, damaged its reputation. According to [27], the actions looked like the situation when you realize the market (or your competitive edge) is going to vanish, so the best move is to "*suck it* *dry, fast*". The comment section of the publication [27] included many good comments but also many strongly biased comments against MDPI (mostly anonymous), see Figure 4.

> MDPI is probably the most predatory publisher. Journals like "Viruses", "Animals" are bullshit. Their portfolio is a portfolio of predatory and awful Journals. Do not count MDPI publications as Publications, but as fake and junk Academic articles. Reject candidates for a position in your faculty if they published in MDPI even 1 paper. Please, spread the Information all over the world thar MDPI is a catastrophe of the Academic system. They are scam, sham, fake and predatory.

**Figure 4.** An example of a bashing comment with an elitist viewpoint suggesting that even a single paper published in a MDPI journal would disqualify the entire career of a researcher.

So, is there something fundamentally wrong with the special issue strategy? Predatory journals market all possible topics in a wide range of research areas with very generic topics so in this regard the action is indeed similar. If the journal fails to identify its target group, aggressive marketing becomes spamming. The other side of the coin is that do you receive enough call-for-papers from reputed journals about their special issues? Do they market enough? And do we *need* these special issues at all?

Let us analyze the question a bit. First, researchers study topics *they* decide; not the topic of the *journal* (or guest editor) wants to. On one hand, a too narrow topic will cause a lack of submissions as it relies mainly on people working in the same niche area. On the other hand, invitations via general mailing lists will annoy many. Marketing the special issue does serve the purpose of marketing the journal as the researchers get to know publication opportunities for their papers.

We will next share our experience when serving as a guest editor for MDPI journals (*Algorithms*, *Applied Sciences*, *Remote Sensing*, *Entropy*) and a Frontiers journal (*Frontiers in Robotics and AI*). The main lesson learned was that it is hard to attract people to submit no matter how broad (or niche) the topic. The best experience was to combine the special issue with a competition. We experienced this twice. Neither of them led to a huge number of submissions, but our knowledge increased significantly on those topics and resulted in two good summary papers [40, 41].

Several negative experiences were collected as well. But not because of the special issue model itself or the way they were organized. On the contrary, the staff gave us full support. Problems surfaced with our own submissions rooted in two factors: poor editorial work, and a lack of coordination between the inviter (journal) and the decision makers (editors). Three times our paper was rejected despite the papers all being valid. Two times the paper was considered *out of scope* (relevance bias) of the special issue showing lack of coordination between the journal and the guest editor. Once the paper was rejected because of the incompetence of the editor to deal with a hostile reviewer; see [**Appendix B; Case 1**].

There are indeed problems with the special issues because MDPI uses it aggressively as a marketing tool. In traditional journals, they are usually managed by academic editors. Crosetto [27] also noted the separation of the roles that, in the case of MDPI, it is the *publisher* who sends out invitations for the special issue and it is unclear if the editorial board has any role in this. Our experience confirms this. The academic editors showed very little interest in the special issue (or the journal in general) or seemed not to know what was going on.

Organizing special issues may take a lot of effort. It can serve the purpose for younger researchers to gain editorial experience as it is useful to see all sides of the table. In our case, the positive experiences came when we aligned the special issue with a competition. Otherwise, we conclude that special issues are mainly a marketing tool. See [**Appendix B**] for more details.

## 6. Discussion

We next discuss two possible improvements for the peer review system inspired by the experiences learned in this paper: different review models, and support of AI.

### 6.1. Different review models

**Open review**: It is expected to reduce the number of garbage reviews if your name is known by the authors. There would be fewer reject recommendations based on secondary criteria, and less elitist behavior. Both these benefits can be achieved without losing quality control if the review report were publicly available to everyone. Or would you accept papers with false results with your name added there as a reviewer? According to Besancon et al. [42], the researchers of the *alt.chi* track on the *Human-computer interaction conference* (CHI) clearly preferred publishing the review reports and the identity of the reviewers. Haffar et al. [4] observed open review has a small favorable effect on the quality.

**Rewards**: It creates motivation and increases the efforts put into the review. Instead of picking a few weak points and writing a pseudo review, the reviewer is more likely to provide some constructive comments even for a weaker paper. Discount in publication fee is used by MDPI. Allen et al. [43] discussed other possibilities, but the challenge is what kind of incentives would have the best effects both on the quality and efficiency of the reviews.

**Efficiency:** A good review does not require months. Nobody is going to spend many hours on a typical paper review. It is mainly a question about allocating time for it and not letting the paper stay idle on the reviewers' desk and then rushed to a quick-and-dirty review report by the last-minute deadline reminder. Rushed reviewers are more likely to provide garbage reviews. In my opinion, the process should be more efficient, involving clear deadlines with polite but persistent reminders. Good editorial work allows flexibility when important.

Quality of review is more important than speed but only if the review reports are provided fast. Three months for the first decision is good in several ways. First, it allows the authors to take a break from the paper and see the content later with fresh eyes. Second, it allows more error fixing and quality improvements. But if it leads to garbage reviews without any feedback on the content, it is a waste of time, and repeating the process many times becomes counter-productive. A very rapid process can also be harmful as it may compromise the quality, but too long a review process is even more harmful. Authors will lose their motivation and focus.

**6.2. Artificial intelligence**

Many parts of the review process can be automated. Artificial intelligence (AI) tools can be developed to automate the key steps: finding reviewers, managing the process, and even making automated reviews to a certain extent. There are already good tools to detect plagiarism and potential reviewer matching to start with. Considering that AI is stepping into every aspect of our lives, it would be surprising if we cannot find ways for AI to help the peer review process as well.

Adopting a machine learning approach as such would just copy all the hidden biases in the training material. For example, Checco [44] pointed out that authors from a country with a historically high rejection rate may share a similar writing style (language transfer), and therefore, become a victim of penalized by the negative bias by the learned AI. His findings were that AI has a high potential to evaluate scope and formatting, and detect plagiarism, but low to evaluate novelty, impact, and soundness.

According to my experience, most submitted papers suffer from poor readability. At the same time, assessing readability is one of the easiest tasks where AI is expected to penetrate fast. Lebrun [45] introduced four principles for the good structure of a paper, and some of these principles have been implemented in an automated tool by Kinnunen et al. [46]. Another automated tool for providing immediate feedback to authors was developed by Harwood [47]. ChatGPT could also turn to an excellent tool for analyzing readability, as it was built as a language model, but unreliable checking factual content as demonstrated by de Grijs [48].

Peer review cannot be completely replaced by automated tools, but they may help to improve the quality of the papers [49], initial quality control and finding reviewers [50]. Recognizing the reliability and validity of the results is more challenging and likely the part where humans will remain in the loop.

7. **Conclusions**

MDPI has been criticized for questionable editorial practices and even labeled as a predatory publisher. We recognized three questionable practices:
- Cultivating self-citations
- Size of editorial board
- Special issue practices

Boosting impact factor by self-citation is typical even for reputed journals. Because of this, impact factor has already lost its status as the primary quality measure; it is just one of the many indirect indicators. According to Havge [6], uncritical use of the impact factor is not the solution, objective quality criteria are still needed, and the focus should be on replacing impact factor with a better and fair quality assessments.

Having a large editorial board is motivated by the large volume of submissions faced by journals these days. It keeps the workload of an individual editor small and allows a better topic match of the paper and the editor. A potential consequence is the lower commitment by the editors. The same effect has been detected in the number of co-authors in a paper: the more authors the less the commitment. According to my experience, the best team size is 3, which has been also empirically observed by Bramoulle and Ductor [51]. More authors degrade the individual commitment and the overall quality of the paper more than the increase of authorships brings in. The same phenomenon is likely to play a role in editorial commitment.

Special issue practice is mainly a marketing tool and is considered spamming if it fails to reach its target audience. It also enhances relevance-bias and can therefore become counter-productive. I stopped paying attention to special issues as they rarely match my research interests.

The problems are in the peer review process. The efficiency and quality of the review process in existing journals require improvements. The suggestions of Oviedo-Garcia [19] to effectively ban MDPI journals as being predatory seems over-reaction. The problems in the peer review have been known for a long time and the review process is considered by many as "*playing the game*". Petrou [15] even speculated that: "*Could it be that the MDPI model of rapid peer review is better at accelerating discovery than preprints or traditional, slow peer review?*"

Different forms of review have been considered but surprisingly little used. Worth considering are open reviews and using rewards like vouchers to compensate the efforts of the reviewers. Reviewer motivation is important: an unmotivated reviewer will just recommend rejection and find easy excuses hoping not to see the paper again. The same paper then continues to circulate and if no real feedback is given, the authors just learn to play the game instead of improving quality. Motivated reviewers are likely to spend more effort and be constructive. Open non-anonymous review should eliminate most garbage reviews.

Enforcing quality is tough as it requires expertise, and so far, human in the loop. Later, AI will gradually replace humans in most parts here, too. Efficiency, however, is not *rocket science* and many parts in this regard can be automated.

Allen et al. [43] suggested using paid professional reviewers which is worth considering. MDPI is partially implementing this already. A good language proof-reading can also improve readability. Vines and Mudditt [52] pointed out several potential negative consequences of paid review including the birth of review factories, motivation shift from professional duty to maximizing profit, and many other side-effects of the increased publication fees especially in open access journals. Implementing this successfully would not be trivial.

To sum up, hunting fake journals (and papers) by too strong measures has a negative effect on the publication world. Banning an entire publisher would reduce the publication opportunities especially for younger researchers which will burden the remaining journals even more. It would cultivate elitism by making publishing more difficult. Journals with high quality standards are still needed but if it is implemented merely by elitist selection with secondary criteria, it is a cream-cropping. Without proper feedback, the remaining papers continue to circulate without any improvement.

To keep the peer review system working, we must have two distinctive goals: encourage research by publishing valid papers and reject flawed papers. The problem is not a single-objective optimization where we reject anything not meeting some secondary criteria. The coin has two sides.

## Conflict of interest

The authors declare that there are no conflicts of interest.

**Author's biography**

**Dr. Pasi Fränti** has over 30 years of experience in research and has been a tenured professor of computer science since 2000. He has published over 120 peer reviewed journal publications, 180 peer reviewed conference publications, 64 other scientific publications, and 1 paper in a predatory journal. He has served as associate editor for Pattern Recognition Letters, Journal of Electronic Imaging, Machine Learning with Applications, and Applied Sciences (MDPI). He has been a guest editor for 6 journals (4 by MDPI). He is one of the founding editors of the AIMS Journal of *Applied Computing & Intelligence*[3] [53], which uses a diamond open access model (free of any charges).

---

[3] https://aimspress.com/journal/aci

## Appendix A. Experiments on inefficiency of journals

**Case 1:** Three years process. Accepted without any review reports.

The paper [24] was in process for three years and eventually accepted without even a single review report. This was in the era when moving from manual to electronic submission systems and the journal struggled in the process. A shorter version was prepared first for a conference [54], and it provided **seven** short but useful review reports that helped to improve the journal version as well. In 2003, the journal messaged they had overcome their "administrative problems" but sent another message 6 months later that they were still facing difficulties in finding reviewers. Eventually, we sent a revised version explaining what changes were made based on the review of the conference version. We also provided a few names of colleagues who agreed to review the paper. After one month, the paper was accepted as such. No review reports were ever provided.

Fortunately, this was not an urgent paper even though the conference presentation initiated an invited publication [55] to GIM International magazine. The method was both "ahead and behind" of its time in terms that everyone expected vector maps to make a breakthrough in mobile devices at that time. Contrary to these expectations, raster maps remained the-state-of-the-art for practical reasons and similar approaches to our dynamic buffering model was adopted into some practical solutions due to its simplicity.

**Summary of the process:** (Blue fonts indicate correspondence by the journal)

25.5.2001: The paper was originally submitted (before the electronic system)

10.6.2001: We have been moving over to electronic submission and review in recent months and it will be very helpful if you are able to let me have a PDF version.

27.2.2002: The results were presented at the Geomatics Symposium and revised according to feedback there.

26.7.2002: There has been a delay in processing your manuscript. We have had some administrative problems in recent months and as a result several manuscripts have been in the system for longer than I would normally like. Unfortunately your manuscript is one of those affected but I am pleased to say the problem has now been rectified.

23.1.2003: I am pleased to report that we have now overcome the administrative difficulties you mention.

22.7.2003: We are still experiencing difficulties in getting referees.

21.4.2004: We have revised the paper by improving presentation (fig.1,4,10 have been added), and related work has been cited [20,21,28,29] covering the file format issues [20], the compression research [21], and the application [28,29] in more detailed. The main architecture, however, is still pending (this paper submitted to IVC journal)."

19.5.2004: Accepted.

**Case 2**: Four years without any real review. Published meanwhile elsewhere.

The paper [25] was original submitted to a journal and acknowledged to have received it but seemed never to progress from the "*with editor*" status to the actual review. The journal did not respond to our query. After four months of idle, we decided to submit it to Pattern Recognition Letters where it was accepted 8 months later. The original submission we left hanging out of curiosity to see how long it would take for the journal to take any action. After 4 years, the journal provided a reject decision claiming that "*half the same*" version was found on the web. It was the preprint version of the published paper on our website. We never submitted to this journal ever again.

**Summary of the process:**

14.5.2010: Submitted to Pattern Analysis and Applications

6.6.2010: "With editor" status.

24.8.2010: Enquiry about the status.

10.9.2010: Submitted to Pattern Recognition Letters

23.6.2011: Accepted.

19.5.2014: Rejected since "half the same" was found on web (preprint of the published paper).

**Case 3**: Four years with garbage reviews

Our recent paper [33] has been submitted to three different IEEE journals (TPAMI, TKDE, TII) and Information Sciences. IEEE is a reputed brand and a natural forum for any good research paper in our field. It is also expected to provide good reviewer reports with useful feedback in 3 months. Our paper is about a simple and general post-processing technique, which can be applied to any existing outlier detector. Experiments showed improvements on 12 existing outlier detectors. However, the reviewers could not tolerate the simplicity of the idea and expected a more complex solution.

The first journal (TPAMI) took only 3 months, but the result was only cream-cropping attitude (*nothing new*). The second journal (TKDE) took 12 months which has been the norm for this journal for most of our submissions. This time there were only two reviewers of which only recommended rejection but with the same, *nothing new*, attitude.

The third journal (TII) took 6+2+1 months in total consisting of two re-submissions and one rebuttal. As earlier, the reviewers provided mostly useless comments. The only difference was that the reviewer recommendations were not shown. One reviewer insisted that the work is not novel and kept on providing literature references time after time and we kept showing, one by one, that those papers merely apply (an existing) outlier detector, and the method used is included in our comparisons already.

IEEE supports that author can write rebuttal if they disagree with the journal decision. This is what we did. The editor-in-chief never even bothered to reply until we contacted to IEEE vice president of publications who took the matter more seriously. He provided detailed feedback on how to make the contribution easier for the readers to digest its contribution. This was the first time we received any useful feedback during the entire review process. The paper was improved, accordingly, but the result did not change. The journal editor re-processed the paper but provided merely a copy-paste of the original decision without any new valid content. The problem was not the lack of clarity, but the method being too simple for the reviewers to digest.

The paper was later submitted to Information Sciences providing rejection with a collection of seven poor-quality garbage reviews. Eventually, we re-submitted the paper once more to TKDE with a fancier over-selling title with added by some superficial theory, and it got accepted. Some improvements were made during this over-lengthy review process but nothing remarkable; the only significant useful feedback was provided by the vice-president. Most reports were garbage reviews and we received lots of those.

**Summary of the process:**

4.9.2019: Rejected after review from IEEE Trans. on Pattern Analysis and Machine Intelligence (**3 months**)

    #1 Reject: Although this paper is well-written with extensive experiments, the novelty is limited for TPAMI;

> #2 Reject: The proposed model is not novel. It is more like a combination of two strategies.

15.4.2021: Rejected from IEEE Trans. Knowledge and Data Engineering after review from (**12 months**)

> #1 Minor: It is quite simple, but the effect is remarkable. Can be applied with any existing outlier detector.
>
> #2 Reject: I vote for rejection. Idea is straight-forward, it lacks technical contribution and depth.

5.12.2021: Rejected from IEEE Trans. on Industrial Informatics. Allow to re-submit (**6 months**)

> No meaningful comments.

21.2.2022: Rejected from IEEE Trans. on Industrial Informatics. Allow to re-submit (**2 months**)

> #1 Reject: The main deficiency of this manuscript is the idea is ordinary.

30.3.2022: Rebuttal to the editor-in-chief. (**no response**)

16.4.2022: Reminder 16.4.2022. (**no response**)

19.4.2022: Rebuttal to IEEE Vice-president of publications

> Provided useful feedback about how to better explain the key contributions.

14.7.2022: Rejected from IEEE Trans. on Industrial Informatics. Allow to re-submit (**1 month**)

> #1: There are still lots of issues in the paper… with many irrelevant comments.
>
> #2: The reviewer thinks this manuscript needs a massive improvement.
>
> #3: The idea put forward in this article is too simple to be published in such a high-level magazine as TII

3.3.2023: Rejected from Information Sciences (**2 months**)

> Seven reviewer comments; almost no useful comments and mostly garbage.

11.11 2023: Accepted by IEEE Trans. Knowledge and Data Engineering by two review rounds (**5 months**)

## Appendix B. Experiments on MDPI special issues

**Case 1: Algorithms** (poor editorial work + hostile reviewer)

I took it seriously and marketed the special issue via my networks and planned also to publish our paper in the same special issue. The idea was to document an existing and widely used clustering benchmark with evaluation guidelines. This paper would probably not have even been written without the special issue activity. It was meant as the first paper in the special issue to guide other authors planning to submit but this never happened.

The result was very disappointing, and the paper was never even accepted by the journal. It was later published elsewhere [36] and has been cited so far extensively (>500 Google Scholar citations by 10$^{th}$ September 2024). It also inspired us to write a follow-up paper [56] about the never-ending interest in k-means initialization. This follow-up paper has also been a big citation magnet and the most downloaded paper in the Pattern Recognition journal so far.

The process with the MDPI special issue caused lots of frustration and even anger towards the journal. Since our plan was accepted by the journal, we assumed the editors were supporting it. Only later did we realize the invitation was made by the MDPI staff and the editors had no interest whatsoever in the special issue. I wonder if the editor ever even read the plan.

As a guest editor, I dealt with all the submissions except our own, and helped the other authors to improve the quality of their papers. The process of our paper was handled by someone incompetent for the job and completely unaware of our plans. The overall process was dis-organized where the left hand did not know what the right hand was doing.

In total, we had four revisions but with a dead end. The bottleneck was a hostile reviewer with non-sensical comments such as "*clustering is an ill-posed problem*", a proposal to "*use LaTex*", and ending up "*I suggest to go for something more interesting*". Most of his other comments were pointless rambling which just became worse and worse towards the end of the process, see Fig. A1. The ill-motivated reviewer had no plan to provide any sensible or constructive comments but found merely more excuses for his rejection recommendation. Simply unethical behavior.

Eventually, we submitted the paper elsewhere, but surprisingly we found the very same hostile reviewer (not sure how, as we originally did not cite any of his papers) and continued his rambling reviews. This time we knew what to expect and pointed out the situation to the editor immediately in our response letter and asked to exclude a named person as a reviewer. Around the same time, the reviewer showed up in the Q&A of our conference presentation of another paper on clustering and kept asking the same irrelevant questions. I now guessed his identity, and publicly pointed out that his actions were harmful and delayed our journal publication. The reviewer backed off and made only one request - to cite his paper. Our paper was accepted one month later without further incidents.

```
                    e.g. manhattan or so - or by e.g. using a kernel trick and map it back
                    to distances or use kernel k-means, or by learning distances from the
data ...
        --> in general it is not surprising that overlapping is a problem - and a study
how
            to address this systematically is not provided in the paper
        --> your results that overlapping may help is only observed but not explained
-
            considering the common sense in this line one may now argue to find a
data representation
            which increases the overlap ??? - to improve the results - if this is true ?!
            - you should provide explainations/evidence when (and potentially why)
this is actually the case
        - (2) number of clusters - is a never ending story and is primary related to your
                personal intrinsic model of data similarity so if N is the number
                of cluster anything between 1-N can be right depending on your
expectations
                (in practice people may assume spherical clusters and then some more
                useful number of clusters may show up). If you have to have many
cluster
                your raise the model complexity (parameters) and yes this makes the
problem
                more challenging ... - so what should I do --> e.g. split the problem into
parts
                ... which is quite complicated - I think something around graph
clustering or hierarchical (clustering)
```

**Figure A1.** Snippet of the rambling review report in which no (relevant or irrelevant) points could be extracted. The formatting also made the reading like detective work trying to find out where exactly is the comment. Would you be able to detect the nature of this reviewer as editor?

**Case 2: Applied Sciences** (good experience)

We co-edited a special issue with my colleague by organizing a segment averaging competition[4]. The specific problem is quite a niche; a seemingly small sub-problem needed as a part of a more complex system. We therefore preferred to have a good existing solution rather than invest our time in developing our own method. This led us to launch the competition where participants were encouraged to also write a paper about their method.

The competition took a lot of effort to organize, but we learned a lot. The special issue did not attract many papers, probably due to its niche topic. Most participants were students, colleagues, or someone we knew in one way or another. Many of them lacked experience (and motivation) in how to write scientific papers but we helped one PhD student by co-authoring his paper (basically writing it on his behalf). We liked the elegance and simplicity of the key ideas in his approach. This was one of the papers published (this time eventless) in the special issue [57]. The lessons learned were later published in a summary paper [40]. This time we had decided to submit it elsewhere.

---

[4] https://cs.uef.fi/sipu/segments/

**Case 3: Remote Sensing** (anonymous editor over-ruling guest editor)

I defined the specs[5] but the scope was provided but the journal. This time I spent much less effort on the special issue compared to Cases 1 and 2. The scope was probably again niche, and it did not attract many submissions. The journal staff eventually offered to extend the deadline motivated by a financial reward (discount on the publication fee) if it raised enough publications. I did not take the offer (too much work for the benefit).

We had also planned to submit our paper as it had a perfect match to the scope, but the paper did not become ready in time. The journal had meanwhile recruited another author from one of the two published papers in our special issue, to run another special issue[6] with almost the same title. We decided to submit our paper to this issue instead.

This time our paper was rejected by an anonymous editor (lack of transparency) because one reviewer insisted in his view that the paper should be completely re-written from the remote sensing point of view. The guest editor was not even consulted in the decision making. Our contribution was to adopt a classical k-means clustering algorithm for image segmentation, and remote sensing was merely a case study application. The reviewer's proposal was meaningless.

The experience again showed how disorganized the system worked – like a machine without any intelligence. Why invite a clustering expert to run a special issue, accept his scope, but then do not respect it later? The guest editor was willing to accept the paper but was overruled by the editor. The conclusion was that the special issues are invited merely for marketing purposes; knowledge and ideas do not matter, and the guest editor's opinions are not even respected. The paper itself was published later in another journal [58].

**Case 4: Entropy** (test of the system)

This was the last time I accepted any guest editorial task from MDPI. I accepted it merely to see how the system would work if I joined as another robot in the system without any intellectual effort. I accepted their topic proposal as such. This time the scope was very wide and basically the same as the topic of the entire journal. I promoted the special issue via my networks and by using one of the two long email lists provided by the journal. I automated the emailing to the (mostly unknown) authors in the list which caused a temporary ban of my email by MS/Outlook systems. Even if I do not use the system myself, there are still other Outlook users who occasionally fail to receive my emails ending up in their spam folder.

After this incident, I stopped any further promotional work for this special issue, and merely followed up the remaining active contacts. Only one paper was ever published in this special issue, and it was even later moved to another issue. We did not have anything useful to submit ourselves at that time. The main conclusion became now even clearer, the guest editorials are used just as a marketing tool by recruiting volunteers work for act as guest editors. For us, it served the purpose when we organized it as a challenge, but this scheme required lots of effort.

---

[5] https://www.mdpi.com/journal/remotesensing/special_issues/RS_Clustering_Algorithms

[6] https://www.mdpi.com/journal/remotesensing/special_issues/CA_Hyperspectral

**Case 5: Frontiers** (is the grass greener on the other side?)

This one was for another publisher (Frontiers) but is worth mentioning as it is a similar open access publisher. This time we also arranged the special issue as a competition. We found the competition useful and learned a lot on the topic. The competition itself attracted only three valid submissions (one of our own), of which only two resulted in a paper. The winner of the competition was the author (Keld Helsgaun) of the state-of-the-art holder of large-scale travelling salesman problems based on his existing LKH method [59]. From the results, we wrote a summary paper on the same issue and consider it a nice one worth reading [41].

Besides the topical knowledge, we also learned how the online review system of Frontiers works with a rapid (almost live) discussion with the reviewers. This is a good alternative and achieves the same efficiency as MDPI. But again, the quality of the editorial still matters a lot as the process can easily end-up with multiple never-ending rounds of revisions that will conquer all your time. Luckily the editor concluded the process fast enough. There were a few downsides though:

(1) benefits were rather limited (if any);
(2) non-responsive to queries during the process;
(3) requirements about the content after acceptance;

The last two remind the dis-organization of MDPI. The main benefits were the lessons learned about the topic which came because we organized the special issue as a competition. Another benefit was getting to know relevant people on the topic. Otherwise, the guest editorial served merely as a CV addition. Unlike MDPI, Frontiers did not provide any significant reduction of the publication fee. They also never replied to our query whether the publication fee would cover language corrections. Since the publication fee was rather high it would have been good to know if we needed also to cover possible language proof-reading.

The most annoying aspect of the operation was a last-minute requirement to reduce the number of Figures in the paper, or to put some of them into a supplement. This happened *after* acceptance (and the fee payment). The staff suggested a possible detour by merging some pictures to reduce their count. Both suggestions were unappealing and would have somewhat degraded the readability of the paper. We disagreed and negotiated successfully over this annoying demand.

**Case 6**: **ISPRS International Journal of Geo-Information** (editor over-ruling guest editor)

This special issue was not organized by us, but it still serves the purpose of showing how disorganized the special issue strategy is. I received an invitation to submit a paper with a 100% discount. The special issue was not an exact match to our research but close enough; the topic of the journal itself was a good match. I wrongly assumed the inviter knew something about my research to offer a full discount.

We submitted a paper that was not hugely significant, but we felt the negative experience of using Medoid in segment averaging needed to be reported. At least we needed a clear citation to literature for this result. I considered it as a training paper for a freshly started PhD student who needed to gain experience. So, we submitted a paper [60].

The problems were about relevance and surfaced only at the final stage. The guest editor was friendly and pointed out the mismatch of topics and asked to add more about our web tools used, making it closer to the special issue theme about Web GIS architectures and applications[7]. We acted accordingly and considered it a good idea. This extension also motivated me to prepare more material along this theme leading to a keynote speech in the DAMSS workshop.[8]

---

[7] https://www.mdpi.com/journal/ijgi/special_issues/web_GIS

However, while the review process went smoothly, the editor (this time by his real name) interrupted the process and rejected the paper as being out of scope. The paper was published later in the same journal but as a normal paper outside of the special issue. While the result was positive, the process itself was unprofessional and not the kind of first impression you want a starting PhD student to get, what is the scientific publishing world like.

©2024 The Author.

---

[8] https://www.mii.lt/damss/index.php/damss/damss-2021